\documentstyle[12pt,thmsa,sw20nhj]{article}

\input{tcilatex}
\begin{document}

\title{}
\author{\and A. Tartaglia \\
Dip. Fisica, Politecnico, Torino, Italy. E-mail: tartaglia@polito.it}
\maketitle
\title{Geometric treatment of the gravitomagnetic clock effect.}

\begin{abstract}
We have developed a general geometric treatment of the GCE valid for any
stationary axisymmetric metric. The method is based on the remark that the
world lines of objects rotating along spacely circular trajectories are in
any case, for those kind of metrics, helices drawn on the flat bidimensional
surface of a cylinder. Applying the obtained formulas to the special cases
of the Kerr and weak field metric for a spinning body, known results for
time delays and synchrony defects are recovered.
\end{abstract}

\section{Introduction}

The gravitomagnetic clock effect (GCE) consists in the loss of synchrony of
identical clocks carried around a massive spinning body, in opposite
directions. This effect is a consequence of general relativity and its
presence has been foreseen in connection with the so called
gravitomagnetism, i.e. that part of the gravitational field which, in weak
field approximation behaves as the magnetic part of the electromagnetic
interaction \cite{mush1}. The GCE has been considered as an interesting an
promising means to test the general relativistic influence of the angular
momentum of a mass on the structure of space time nearby, and in particular
on the pace of clocks orbiting around the body \cite{tart1}\cite{mush2}.

Actually the GCE is strictly akin to the Sagnac effect, which is a special
relativistic effect induced by pure rotations, first considered as a purely
classical effect by G. Sagnac \cite{sagnac}, further on recognized in its
real nature and studied by several author (see for instance \cite{anandan} 
\cite{rt}).

A relevant aspect both of GCE and of Sagnac effect is their genuine
geometrical character. The present paper moves from the remark that the
world line of an object steadily moving along a spacially circular
trajectory around a symmetry axis of the gravitational field is a helix, to
derive a general method for describing the GCE from different viewpoints in
general terms, without being a priori limited to the Kerr metric or to its
weak field limit as in \cite{mush1}, \cite{tart2} and \cite{tart1}. Finally
the general method, when applied to the mentioned special cases, will
reproduce the known results, but will allow also for an easier recognition
of the situations more viable for experimentally detecting time delays and
synchrony defects.

\section{General geometric features of the gravitomagnetic clock effect}

As already said in the introduction the space time of a massive body
steadily rotating about an axis at rest with respect to distant galaxies has
a simple and interesting property: the world line of any object orbiting the
central mass at a constant distance from the axis is a helix.

In other words, for any such object a cylinder (bidimensional surface)
exists on which the world line is drawn; this surface is flat and when
opened in a plane the world line becomes a simple straight line. This is
true irrespective of the global curvature of space time along the
fourdimensional trajectory.

Viewing things this way allows one to catch at a glance both the Sagnac and
the GCE. Consider the $1+1$ cylinder opened in a plane, shown on figure 1.

\FRAME{ftbpFU}{4.107in}{3.0839in}{0pt}{\Qcb{The figure shows the development
on a plane of the world tube on which the world lines of steadily rotating
objects are drawn. The helices are now straight lines; points at $\pi $ are
identified with points at $-\pi $ . Event $C$ corresponds with the
conjunction (after a whole revolution) of two oppositely rotating objects; $A
$ and $B$ are the respective completions of a revolution as seen by a
''fixed star'' observer; $D$ and $E$ are the equivalent events seen by a
rotating observer.}}{}{disegnogrg11.gif}{
}

Oblique lines are the developed helices; on the horizontal axis angular
coordinates are reported, the vertical one shows a variable proportional to
the inertial time of an observer at rest with the symmetry axis; points at $%
\phi =\pi $ are identified with points at $\phi =-\pi $ and the same $t$\cite
{rt}. Objects steadily rotating with different speeds are represented by
different slope lines; it is evident that such lines cannot cross each other
on the vertical of the origin (vertical dotted line on O) unless they are
symmetric with respect to it, i. e. unless the speeds are the same in
magnitude but oppositely directed. Similarly it is evident that also the
intervals between two successive crosspoints (two conjunctions) are
different if measured along the two lines, unless the slopes are symmetric:
the proper times between conjunctions are different.

These simple facts are indeed the graphic explanation of the Sagnac effect:
a rotating observer is in his turn represented by a helix in space time
(straight line on fig. 1), whose interceptions with the world lines of the
test bodies determine proper intervals and proper time differences typical
of the effect. The situation is not different when a gravitational field is
present, provided it possesses an axial symmetry and no angular momentum as
it is the case for a Schwarzschild metric. Simply the gravitational field
will affect the conversion between the proper time of the rotating probes
and the coordinate time, but all the typical Sagnac phase effects will be
there.

When an angular momentum must be accounted for, the scheme remains in
principle the same, but a careful discussion of the viewpoint of different
observers needs be made.

\subsection{Graphic representation of the non zero angular momentum case}

When the gravity source possesses an angular momentum two objects rotating
in opposite directions with the same coordinate speed have again a
conjunction at the same coordinate time and the same proper time. However
two freely orbiting objects on the same circular trajectory and opposite
directions do so in general at different angular coordinate speeds, have
conjunctions at different proper times and cross the line of sight of a
distant inertial observer at different coordinate times. The image given in
fig. 1 may be retained but for the fact that now the axes of $t$'s and $\phi 
$'s are no longer orthogonal: the surface on which the world lines of
circularly rotating objects are drawn, are still flat and the plane
representation of the situation is shown on fig. 2.

\FRAME{ftbpFU}{4.107in}{3.0822in}{0pt}{\Qcb{This scheme is the equivalent of
fig. 1, but for the fact that now an angular momentum is present and
consequently the $t$ and $\phi $ axes are no more perpendicular.}}{}{%
disegno2.gif}{
}

Everything may be described by the use of simple methods of bidimensional
Minkowskian geometry. The reference frame is the one drawn in the figure.
The world line of a rotating object may still be written as 
\begin{equation}
t=\frac{\phi }{\omega }  \label{retta}
\end{equation}
provided it passes through the origin event. We assume that when $\omega >0$
the object is corotating with the source of the gravitational field; the
reverse when $\omega <0$.

Considering the rotation symmetry, we must complement (\ref{retta}) with the
condition that, when the running event reaches the borders of fig. 1 or fig.
2 the representative angle bounces back ($\omega >0$) or ahead ($\omega <0$)
by a $2\pi $ term. In other words the world line becomes 
\begin{equation}
t=\frac{\phi \pm 2\pi }{\omega }  \label{wl}
\end{equation}
where the $+$ sign corresponds to corotation and the $-$ one to
counter-rotation. After one more turn an additional $2\pi $ is introduced,
and so on.

Equipped with these simple definitions and rules we immediately see that the
coordinate time for a complete revolution (distant observer view point)
which is of course $T=\frac{2\pi }{\left| \omega \right| }$ corresponds to
the proper time interval of the revolving object 
\begin{eqnarray}
\tau &=&\frac{1}{c}\sqrt{g_{tt}T^{2}+2g_{t\phi }2\pi T+g_{\phi \phi }4\pi
^{2}}  \label{proprio} \\
&=&\frac{2\pi }{c\left| \omega \right| }\sqrt{g_{tt}+2g_{t\phi }\omega
+g_{\phi \phi }\omega ^{2}}  \nonumber
\end{eqnarray}
In flat space time this would be $\tau =T\sqrt{1-\beta ^{2}}$; in the
general case the presence of a term containing $g_{t\phi }$ accounts for the
non orthogonality of the reference axes (polar coordinates are understood).

It is useful to work out the position of the first intersection event
(conjunction) of two objects endowed with different angular velocities $%
\omega _{1}$and $\omega _{2}$; let us assume by default that $\omega _{1}>0$
and $\omega _{1}>\omega _{2}$. If it is also $\omega _{2}>0$ the first
conjunction is found when the equation 
\[
\frac{\phi +2\pi }{\omega _{1}}=\frac{\phi }{\omega _{2}} 
\]
is satisfied, i.e. when 
\[
\left\{ 
\begin{array}{l}
\phi =\phi _{a}=2\pi \frac{\omega _{2}}{\omega _{1}-\omega _{2}} \\ 
t=t_{a}=\frac{2\pi }{\omega _{1}-\omega _{2}}
\end{array}
\right. 
\]
The proper times of the two objects at the conjunction are 
\[
\tau _{1,2}=\frac{2\pi }{c\left( \omega _{1}-\omega _{2}\right) }\sqrt{%
g_{tt}+2g_{t\phi }\omega _{1,2}+g_{\phi \phi }\omega _{1,2}^{2}} 
\]
corresponding to a synchrony defect 
\begin{eqnarray}
\delta \tau _{12a} &=&\left( \tau _{1}-\tau _{2}\right) _{a}  \nonumber \\
&=&\frac{2\pi }{c\left( \omega _{1}-\omega _{2}\right) }\left( \sqrt{%
g_{tt}+2g_{t\phi }\omega _{1}+g_{\phi \phi }\omega _{1}^{2}}-\sqrt{%
g_{tt}+2g_{t\phi }\omega _{2}+g_{\phi \phi }\omega _{2}^{2}}\right)
\label{difetto1}
\end{eqnarray}

If it is $\omega _{2}<0$ and $-\omega _{2}>\omega _{1}$the values are 
\begin{equation}
\left\{ 
\begin{array}{l}
\phi _{b}=2\pi \frac{\omega _{1}}{\omega _{1}-\omega _{2}} \\ 
t_{b}=t_{a} \\ 
\delta \tau _{12b}=\delta \tau _{12a}
\end{array}
\right.  \label{difetto2}
\end{equation}
Finally, when $\omega _{2}<0$ and it is $-\omega _{2}\leq \omega _{1}$%
\begin{equation}
\left\{ 
\begin{array}{l}
\phi _{c}=\allowbreak 2\pi \frac{\omega _{1}+\omega _{2}}{\omega _{1}-\omega
_{2}}=\phi _{a}+\phi _{b} \\ 
t_{c}=\allowbreak \frac{4\pi }{\omega _{1}-\omega _{2}}=2t_{a} \\ 
\delta \tau _{12c}=2\delta \tau _{12a}
\end{array}
\right.  \nonumber
\end{equation}

A relevant situation is that of circular geodetic motion at a constant
coordinate radius $r$. To study this case let us start from a metric whose
non zero elements are $g_{tt}$, $g_{t\phi }$, $g_{rr}$, $g_{r\theta }$, $%
g_{\theta \theta }$ and $g_{\phi \phi }$; all of these elements depend on $r$
and $\theta $ only. Imposing the conditions $r=$ constant, $\theta =$
constant $=\pi /2$ with a symmetry such that all the metric elements are
extremal for the chosen $\theta $ value, the equations of geodesic motion
lead to the expression 
\begin{equation}
g_{\phi \phi ,r}\omega ^{2}+2g_{t\phi ,r}\omega +g_{tt,r}=0
\label{geodetica}
\end{equation}
Commas mean partial differentiation with respect to the variable after them.

Angular velocities of (spacely) circular geodesic motion are then 
\begin{equation}
\omega _{\pm }=\frac{-g_{t\phi ,r}\pm \sqrt{g_{t\phi ,r}^{2}-g_{\phi \phi
,r}g_{tt,r}}}{g_{\phi \phi ,r}}  \label{omega}
\end{equation}
This can be written 
\begin{equation}
\omega _{1,2}=\omega _{0}\pm \omega _{*}  \label{om1}
\end{equation}
where $\omega _{0}=-g_{t\phi ,r}/g_{\phi \phi ,r}$ and $\omega _{*}=\sqrt{%
\omega _{0}^{2}-g_{tt,r}/g_{\phi \phi ,r}}$.

Using (\ref{om1}) and arranging summations so that $\omega _{1}>\omega _{2}$%
, we can calculate the synchrony defect at conjunction for two freely
counter-orbiting objects.

To end this section let us still consider the situation as viewed by an
observer who rotates with an angular velocity $\Omega $ of his own. In the
proper time of this observer the revolution period of a prograde orbiting
object is deduced from (\ref{difetto1}) with $\omega _{1}=\omega _{0}+\omega
_{*}$ and $\omega _{2}=\Omega $ ($\omega _{1}>\Omega >0$), obtaining 
\begin{eqnarray*}
\tau _{+} &=&\frac{2\pi }{c\left( \omega _{1}-\Omega \right) }\sqrt{%
g_{tt}+2g_{t\phi }\Omega +g_{\phi \phi }\Omega ^{2}} \\
&=&\frac{2\pi }{c\left( \omega _{0}+\omega _{*}-\Omega \right) }\sqrt{%
g_{tt}+2g_{t\phi }\Omega +g_{\phi \phi }\Omega ^{2}}
\end{eqnarray*}
The retrograde case (now $\omega _{1}=\Omega $ and $\omega _{2}=\omega
_{0}-\omega _{*})$ corresponds to a revolution time 
\begin{eqnarray*}
\tau _{-} &=&\frac{2\pi }{c\left( \Omega -\omega _{2}\right) }\sqrt{%
g_{tt}+2g_{t\phi }\Omega +g_{\phi \phi }\Omega ^{2}} \\
&=&\frac{2\pi }{c\left( \Omega +\omega _{*}-\omega _{0}\right) }\sqrt{%
g_{tt}+2g_{t\phi }\Omega +g_{\phi \phi }\Omega ^{2}}
\end{eqnarray*}
Then the proper time difference between the conjunctions with the observer
will be 
\begin{eqnarray}
\delta \tau &=&\frac{2\pi }{c}\frac{\left( 2\Omega -\omega _{1}-\omega
_{2}\right) }{\left( \omega _{1}-\Omega \right) \left( \Omega -\omega
_{2}\right) }\sqrt{g_{tt}+2g_{t\phi }\Omega +g_{\phi \phi }\Omega ^{2}}
\label{diffe} \\
&=&\frac{4\pi }{c}\frac{\Omega -\omega _{0}}{\omega _{*}^{2}-\left( \Omega
-\omega _{0}\right) ^{2}}\sqrt{g_{tt}+2g_{t\phi }\Omega +g_{\phi \phi
}\Omega ^{2}}  \nonumber
\end{eqnarray}
The same quantity, expressed in terms of coordinate times, is 
\[
\delta t=2\pi \left( \frac{1}{\omega _{0}+\omega _{*}-\Omega }-\frac{1}{%
\Omega +\omega _{*}-\omega _{0}}\right) =\frac{c\delta \tau }{\sqrt{%
g_{tt}+2g_{t\phi }\Omega +g_{\phi \phi }\Omega ^{2}}} 
\]

Finally it must be remarked that an inertial distant observer finds also a
difference in revolution times between pairs of freely counter-rotating
objects. One obtains 
\begin{equation}
\delta T=2\pi \left( \frac{1}{\omega _{0}+\omega _{*}}-\frac{1}{\omega
_{0}-\omega _{*}}\right) =4\pi \frac{\omega _{*}}{\omega _{*}^{2}-\omega
_{0}^{2}}  \label{deltap}
\end{equation}

An interesting category of observers are the so called locally non rotating
observers (LNRO) or Bardeen observers . An LNRO is an observer who does not
rotate with respect to matter radially falling towards the central mass;
when the latter is spinning it drags, in a sense, the space time around it
(though the image of a ''drag'' is not really appropriate as pointed out in 
\cite{mush2}), so that a locally ''non rotating'' observer is actually seen
as rotating from another inertial far away observer (distant stars); its
angular velocity is $\Omega _{LNRO}=-g_{t\phi }/g_{\phi \phi }$ \cite
{straumann} and its motion is in general non geodesic. In fig. 2 $\Omega
_{LNRO}$ is a measure of the slope of the cylinder.

Another situation that could be of importance for experimentation is the one
of a rotating observer (angular speed $\Omega $) who sends with opposite but
locally equal velocities (in the tangent space) two objects along his own
path. If ${\bf u}_{o}$, ${\bf u}_{1}$ and ${\bf u}_{2}$ are the
fourvelocities of the observer and the two objects the condition for the
equality of the velocities with respect to the observer is 
\begin{equation}
{\bf u}_{o}\cdot {\bf u}_{1}={\bf u}_{o}\cdot {\bf u}_{2}  \label{scalare}
\end{equation}
Using (\ref{retta}) and the normalization condition for the fourvelocity of
the observer (\ref{scalare}) transforms into 
\[
\frac{g_{tt}+2g_{t\phi }\omega _{2}+g_{\phi \phi }\omega _{2}^{2}}{%
g_{tt}+2g_{t\phi }\omega _{1}+g_{\phi \phi }\omega _{1}^{2}}=\left( \frac{%
g_{tt}+\left( g_{t\phi }+g_{\phi \phi \Omega }\right) \omega _{2}+g_{t\phi
}\Omega }{g_{tt}+\left( g_{t\phi }+g_{\phi \phi \Omega }\right) \omega
_{1}+g_{t\phi }\Omega }\right) ^{2} 
\]
Solving for $\omega _{2}$ one finds, besides the trivial solution $\omega
_{2}=\omega _{1}$, the relevant result 
\begin{equation}
\omega _{2}=-\frac{g_{tt}\omega _{1}-2g_{tt}\Omega -g_{\phi \phi }\Omega
^{2}\omega _{1}-2g_{t\phi }\Omega ^{2}}{g_{tt}+2g_{\phi \phi }\Omega \omega
_{1}-g_{\phi \phi }\Omega ^{2}+2g_{t\phi }\omega _{1}}  \label{omega2}
\end{equation}
Combining (\ref{omega2}) with (\ref{diffe}) it is possible to find the time
delay registered by the observer at the passing by him of the two apparently
equal velocity objects. The result is 
\begin{equation}
\delta \tau =-\frac{4\pi }{c}\frac{g_{\phi \phi }\Omega +g_{t\phi }}{\sqrt{%
g_{\phi \phi }\Omega ^{2}+2\Omega g_{t\phi }+g_{tt}}}  \label{deltau}
\end{equation}
As it can be seen $\delta \tau $ does not depend on $\omega _{1}$ i.e. it is
independent from the actual velocity of the objects with respect to the
observer. In Minkowski space time (\ref{deltau}) reproduces the formula of
the Sagnac effect.

\section{Special cases}

The time delays and synchrony defects determined in the preceding section
may be specialized to various different metric tensors. Two cases are
particularly of interest either in principle or for practical reasons: the
Kerr metric and the weak field approximation of the metric of a spinning
object. The relevant metric elements ($\theta =\pi /2$) are: 
\[
\begin{array}{lll}
\text{Metric elements} & \text{Kerr} & \text{Weak field} \\ 
g_{tt} & c^{2}\left( 1-2G\frac{M}{c^{2}r}\right) & c^{2}\left( 1-2G\frac{M}{%
c^{2}r}\right) \\ 
g_{\phi \phi } & -2a^{2}G\frac{M}{c^{2}r}-r^{2}-a^{2} & -r^{2} \\ 
g_{t\phi } & 2aG\frac{M}{cr} & 2aG\frac{M}{cr}
\end{array}
\]

Introducing these expressions into the formulas of the preceding section we
obtain 
\begin{eqnarray}
\omega _{0K} &=&\frac{aGMc}{a^{2}GM-c^{2}r^{3}}  \nonumber \\
\omega _{0wf} &=&-aG\frac{M}{cr^{3}}  \label{zeri}
\end{eqnarray}
and 
\begin{equation}
\left\{ 
\begin{array}{l}
\omega _{*K}=\allowbreak c\frac{\sqrt{GMc^{2}r^{3}}}{a^{2}GM-c^{2}r^{3}} \\ 
\omega _{*wf}=\sqrt{\frac{GM}{r^{3}}\left( 1+\frac{a^{2}GM}{c^{2}r^{3}}%
\right) }
\end{array}
\right.  \label{starri}
\end{equation}

Combining these formulas we obtain 
\begin{equation}
\left\{ 
\begin{array}{l}
\omega _{1,2K}=\frac{c}{a\pm c\sqrt{\frac{r^{3}}{GM}}} \\ 
\omega _{1,2wf}=\sqrt{\frac{GM}{r^{3}}}\left( \pm 1-\frac{a}{c}\sqrt{\frac{GM%
}{r^{3}}}\right)
\end{array}
\right.  \label{tutti}
\end{equation}

It is clearly $-\omega _{2}>\omega _{1}$ consequently the synchrony defect
between two counter-orbiting objects is obtained from (\ref{difetto2}) and (%
\ref{difetto1}). Let us directly calculate the result in weak field
approximation, keeping the first order (in $a$ and $GM/c^{2}$) terms only: 
\begin{equation}
\delta \tau _{12}\simeq 6\pi \frac{GM}{c^{2}r}\frac{a}{c}  \label{circa}
\end{equation}

From the view point of a distant inertial observer the difference in
revolution times for the two counter-orbiting objects is obtained from (\ref
{deltap}) and (\ref{tutti}): 
\begin{equation}
\delta T=4\pi \frac{a}{c}  \label{esatto}
\end{equation}
This exact result (in Kerr geometry) is remarkably independent both from the 
$r$ parameter of the orbit and from the gravity constant $G$. Actually in
weak field approximation and for a spherical homogeneous mass it turns also
to be independent from the very mass of the central object since then it is: 
$a=2R^{2}\Omega _{0}/5c$ ($R$ is the radius of the body and $\Omega _{0}$ is
its rotation speed).

From (\ref{proprio}) we see that the readings of clocks attached to our two
objects after what is seen as a complete revolution by the distant observer
are 
\[
\tau _{1,2}=\frac{2\pi }{c\omega _{1,2}}\sqrt{g_{tt}+2g_{t\phi }\omega
_{1,2}+g_{\phi \phi }\omega _{1,2}^{2}} 
\]
The difference between these readings corresponds, in weak field
approximation, to (\ref{esatto}): $\tau _{1}-\tau _{2}\simeq 4\pi \frac{a}{c}
$. It is also the value that would be found, with the same approximation, by
a LNRO.

In WFA formula (\ref{deltau}) becomes 
\begin{eqnarray}
\delta \tau &\simeq &\frac{4\pi }{c^{2}}\frac{r^{2}\Omega -2aG\frac{M}{cr}}{%
\sqrt{1-2G\frac{M}{c^{2}r}-\frac{r^{2}\Omega ^{2}}{c^{2}}+4\Omega \frac{a}{c}%
G\frac{M}{c^{2}r}}}  \nonumber \\
&\simeq &\frac{4\pi }{c^{2}}r^{2}\Omega \left( 1+G\frac{M}{c^{2}r}+\frac{1}{2%
}\frac{r^{2}\Omega ^{2}}{c^{2}}\right) -8\pi \frac{a}{c}\frac{GM}{c^{2}r}
\label{terra}
\end{eqnarray}
There is a contribution to $\delta \tau $ depending on $a$ but not on $%
\Omega $ and reproducing the result obtained as first order relativistic
correction to the Sagnac effect \cite{tart2}.

\section{Conclusion}

We have shown how a simple geometric vision of the world lines of steadily
rotating objects in axisymmetric metrics endowed with angular momentum
allows for a description and explanation of the GCE. The method evidences
and recovers some interesting results that could lead to experimental
verifications. Using satellites on circular trajectories one would find a
synchrony defect between counter-orbiting identical clocks given (in WFA) by
(\ref{circa}), wich in the case of Earth is $\sim 10^{-16}$s. A much bigger
effect is seen considering revolution times with respect to a fixed
direction in space (with respect to fixed stars); in that case the two
rotation directions correspond to differences in period length given exactly
by (\ref{esatto}) which for the Earth is $\sim 10^{-7}$s.

Another interesting possibility would be to work with a satellite (the
observer) sending light signals in opposite directions along (non geodesic)
closed paths; the relevant formula in this case would be (\ref{terra}) with $%
\Omega $ given by (\ref{tutti}), which produces 
\begin{equation}
\delta \tau \simeq \pm 4\frac{\pi }{c^{2}}r^{2}\sqrt{G\frac{M}{r^{3}}}-12\pi 
\frac{a}{c}\frac{GM}{c^{2}r}  \label{numeri}
\end{equation}
The upper (lower) sign corresponds to a prograde (retrograde) orbiting
observer. The first term in (\ref{numeri}) is the Sagnac effect, the second
one is the correction induced by the angular momentum of the source of
gravity. Again, in the case of the Earth, the correction is in the order of $%
\sim 10^{-16}$s (a hundredth of a period for visible light).

Finally, in case of experiments on the surface of the Earth (non geodesic
equatorial observer, equal speed objects/signals in opposite directions) the
formula is again (\ref{terra}) with $\Omega $ coinciding with the angular
speed of the Earth $\Omega _{0}$. Using the expression of $a$ appropriate
for this case, the formula reads 
\[
\delta \tau \simeq \frac{4\pi }{c^{2}}R^{2}\Omega _{0}\left( 1+\frac{1}{5}G%
\frac{M}{c^{2}R}+\frac{1}{2}\frac{R^{2}\Omega _{0}^{2}}{c^{2}}\right) 
\]
The ''correction'' originated by the angular momentum of the planet is still
in the order of $\sim 10^{-16}$s.

\end{document}